\magnification=1200 \baselineskip=13pt \hsize=16.5 true cm \vsize=20 true cm
\def\parG{\vskip 10pt} \font\bbold=cmbx10 scaled\magstep2

\centerline{{\it Braz. J. Phys.} {\bf 30}, 766 (2000)}\vskip 20pt

\centerline{\bbold Broad Histogram: Tests for a Simple and}\vskip 3pt
\centerline{\bbold Efficient Microcanonical Simulator}\parG
\centerline{Paulo Murilo Castro de Oliveira}\parG
Instituto de F\'\i sica, Universidade Federal Fluminense\par
av. Litor\^anea s/n, Boa Viagem, Niter\'oi RJ, Brazil 24210-340\par
e-mail PMCO @ IF.UFF.BR\par

\vskip 0.5cm\leftskip=1cm\rightskip=1cm 

{\bf Abstract}\parG

        The Broad Histogram Method (BHM) allows one to determine the
energy degeneracy $g(E)$, i.e. the energy spectrum of a given system, from
the knowledge of the microcanonical averages $<N^{\rm up}(E)>$ and
$<N^{\rm dn}(E)>$ of two macroscopic quantities $N^{\rm up}$ and $N^{\rm
dn}$ defined within the method. The fundamental BHM equation relating
$g(E)$ to the quoted averages is exact and completely general for any
conceivable system. Thus, the only possible source of numerical
inaccuracies resides on the measurement of the averages themselves.

	In this text, we introduce a Monte Carlo recipe to measure
microcanonical averages. In order to test its performance, we applied it
to the Ising ferromagnet on a $32 \times 32$ square lattice. The exact
values of $g(E)$ are known up to this lattice size, thus it is a good
standard to compare our numerical results with. Measuring the deviations
relative to the exactly known values, we verified a decay proportional to
$1/\sqrt{counts}$, by increasing the counter ($counts$) of averaged
samples over at least 6 decades. That is why we believe this
microcanonical simulator presents no bias besides the normal statistical
fluctuations. For $counts \approx 10^{10}$, we measured relative
deviations near $10^{-5}$ for both $g(E)$ and the specific heat peak,
obtained through BHM relation.\par

\leftskip=0pt\rightskip=0pt\parG


\vfill\eject

	Monte Carlo methods are applied to statistical physics in order to
measure the thermal average

$$<Q>_T\,\, = \, {\sum_S Q_S \exp(-E_S/T) \over \sum_S
\exp(-E_S/T)}
\eqno(1)$$

\noindent of some macroscopic quantity $Q$ (magnetisation, density, etc).
The temperature $T$ is fixed, and the Boltzmann constant is set to unity.
Both sums run over all possible microstates $S$ available for the system.
The energy (quantity $Q$) corresponding to $S$ is denoted by $E_S$
($Q_S$). Within traditional computer simulations, instead of taking all
possible states, one takes only a finite set of them, i.e. a Markovian
chain of states randomly tossed according to probabilities dictated by the
Boltzmann exponential factors $\exp(-E_S/T)$, the so-called importance
sampling. Thus, one needs to fix a particular value for the temperature
$T$, before running the computer job. In order to determine the full
dependence of $<Q>_T$ upon $T$, one needs to run the job again and again,
for different values of $T$.

        An alternative is to re-write the same average as

$$<Q>_T\,\, = \, {\sum_E <Q(E)> g(E) \exp(-E/T) \over \sum_E g(E)
\exp(-E/T)}\,\,\,\, ,
\eqno(2)$$

\noindent where the degeneracy $g(E)$ counts the number of states with
energy $E$, and both sums run over all possible energies. The
microcanonical average

$$<Q(E)>\,\, = \, {\sum_{S[E]} Q_S \over g(E)}
\eqno(3)$$

\noindent of the same quantity $Q$ corresponds to a fixed energy, i.e. the
sum in (3) runs only over the states $S[E]$ belonging to energy level $E$.
This microcanonical average (3) is simpler than the canonical counterpart
(1) or (2), prescribing exactly the same weight to all averaging states,
i.e. it is a {\bf uniform} averaging process within each energy level {\bf
separately}.

	Only the Boltzmann factor $\exp(-E/T)$ appearing in equation (2)
depends on the temperatute, carrying all thermodynamic information about
the environment which continuously exchanges energy with the system under
study. Contrary to this, both $g(E)$ and $<Q(E)>$ are independent of the
particular way this energy exchange occurs, independent of the
environment: they are more fundamental properties of the system alone,
defined only by its energy spectrum. They are not thermodynamic
quantities, and do not depend upon thermodynamic concerns like
temperature, equilibrium, etc. In practical terms, equation (2) allows one
to determine $<Q>_T$ for {\bf any} value of $T$, from the knowledge of the
energy functions $g(E)$ and $<Q(E)>$.

	The Broad Histogram Method (BHM) [1] relates $g(E)$ with the
microcanonical averages of two macroscopic quantities $N^{\rm up}$ and
$N^{\rm dn}$, measured at the current state $S$. First, one needs to adopt
some protocol of allowed movements which could be performed on $S$,
leading to another possible state $S'$. One can adopt any such a protocol,
the only restriction being its reversibility (if $S \to S'$ is allowed, so
is $S'\to S$). Considering the Ising model, for instance, the protocol
could be chosen to be the whole set of single-spin flips (among many other
alternative choices). Given such a protocol, $N^{\rm up}(S)$ counts the
number of allowed movements which could be performed on $S$, leading to a
fixed energy increment $\Delta{E}$ (which must be chosen a priori,
although the method is also independent of this choice). Analogously,
$N^{\rm dn}(S)$ counts the number of allowed movements decreasing the
energy of $S$ by the same fixed amount $\Delta{E}$. The fundamental BHM
relation [1] is

$$g(E) <N^{\rm up}(E)> \,\, = \, g(E+\Delta{E}) <N^{\rm dn}(E+\Delta{E})>
\,\,\,\, ,
\eqno(4)$$

\noindent where the microcanonical averages of $N^{\rm up}$ and $N^{\rm
dn}$ are defined by equation (3). By knowing these energy functions,
equation (4) allows one to determine $g(E)$ along all the energy axis, in
steps of $\Delta{E}$ --- a constant, unimportant pre-factor is cancelled
out by performing the average in equation (2). Relation (4) is shown to be
valid for any energy spectrum, under completely general grounds [2]. Also,
the same reasoning could be applied for another basic quantity $q$,
instead of the energy, by considering the degeneracies $g(q)$ instead of
$g(E)$. For simplicity, we will restrict ourselves to the case of the
energy.

	Thus, the Broad Histogram Method consists in: $i$) to choose some
protocol of allowed movements, as well as an energy jump $\Delta{E}$;
$ii$) to measure, {\bf by any means}, the microcanonical average $<Q(E)>$
as a function of the energy, as well as $<N^{\rm up}(E)>$ and $<N^{\rm
dn}(E)>$ which determine $g(E)$ through equation (4); $iii$) to obtain the
desired thermal average through equation (2). There is no approximation at
all, and the final numerical accuracy depends exclusively upon the
microcanonical measuring strategy adopted in step $ii$. The method is
reviewed in [3]. Some references where it is used are [4-16].

	Let's briefly analyse hereafter some possible computer strategies
one can adopt in order to measure the microcanonical averages $<Q(E)>$,
$<N^{\rm up}(E)>$ and $<N^{\rm dn}(E)>$, as functions of the energy. A
Markovian chain of states is obtained by performing random movements
transforming the current state into another. These movements are tossed
among some previously defined set of possibilities, another protocol which
has nothing to do with the BHM protocol of {\bf virtual} movements. Both
protocols could even be chosen to be the same, but not necessarily.

	The first direct strategy is to keep always the same fixed value
$E$: by starting from some state corresponding to the desired energy, one
simply rejects any tossed movement which changes the current energy. After
one has already a large enough number of visited states inside this
particular energy level, the same process is repeated for other levels.
Depending on the adopted protocol of (real) movements, this strategy could
lead to ergodicity problems: due to its high rejection rate, one risks to
sample only a biased sub-set of states belonging to energy level $E$,
violating the required uniform visitation. Moreover, it is also an
inefficient strategy, in what concerns the computer time, again due to its
high rejection rate.

	The opposed alternative is to accept also movements leading to
energy jumps, storing separated averages for each energy level, in
parallel. Obviously, this option is much more efficient in what concerns
the computer time. However, one cannot simply accept any tossed movement:
energy increments would occur more often than decrements, because $g(E)$
is normally a fast increasing function of the energy. As a result, at the
end, only states corresponding to the region near the maximum of $g(E)$
would be sampled. Thus, some movement-rejection prescription must be
adopted, and we get back to the uniformity violation problem. One can
adopt some already well established movement-rejection prescription based
on detailed balance arguments. For instance, canonical, fixed-temperature
dynamics could be adopted [4,6], sampling states inside the narrow energy
window corresponding to the fixed value of $T$. In order to get results on
a broader energy window, one can simply superimpose the histograms
obtained for different computer runs corresponding to different values of
$T$.

	Another possibility is to adopt one of the many multicanonical
dynamics [7,9,11,13,14]. Within this approach, one tunes the $E$-dependent
rejection rate during the computer run, in order to obtain the same visit
probability for all energy levels, i.e. a flat distribution along the
energy axis, at the end. In this case, the visitation probability to each
particular state would be proportional to $1/g(E)$. The so-called
multicanonical methods [17-19] are based just on this feature: by
recording the acceptance probability for energy-increasing movements $E
\to E'$, accumulated during the run and which must be equal to
$g(E)/g(E')$ at the end, one gets the function $g(E)$ except for an
unimportant global factor which cancels in equation (2). BHM is completely
distinct from multicanonical approaches in many features. In particular,
the infomations extracted from each $E$-state $S$ belonging to the
averaging Markovian chain are the {\bf macroscopic} values of $N^{\rm
up}(S)$ and $N^{\rm dn}(S)$, not the mere one-more-visit upgrade $V(E) \to
V(E)+1$. That is why BHM gives more accurate results than multicanonical
approaches, even taking into account the same Markovian chain of averaging
states, as shown in [15] where the multicanonical dynamics [18] is adopted
in order to measure $<N^{\rm up}(E)>$ and $<N^{\rm dn}(E)>$. At the end,
$g(E)$ is determined twice, by following the multicanonical traditional
way or, alternatively, by the BHM equation (4), both using data taken from
the same set of averaging states. A clear accuracy advantage for BHM is
reported [15]. Moreover, due to the macroscopic character of the BHM
quantities $N^{\rm up}$ and $N^{\rm dn}$ this advantage still increases
for larger and larger systems.

	Another crucial difference relative to multicanonical approaches
is that BHM requires only the uniformity of visits among the states {\bf
inside each energy level, separately}, in order to get the correct
microcanonical averages. BHM does not need any detailed balance between
visits to different levels $E$ and $E'$, nor the multicanonical flat
distribution along the energy axis. Any dynamic strategy which is good for
multicanonical methods will be also good for BHM (besides the accuracy
advantage quoted in the last paragraph), but the reverse is not true.
Thus, within BHM, other not-so-restricted dynamic strategies could be
used.

	Profiting from this feature, we decided to test a very simple
dynamic strategy inspired by reference [20] (although within a different
method, the dynamic rule introduced in [20] is essentially the same as
presented hereafter). The idea is to avoid movement rejections at all,
within the energy level currently being sampled for averaging purposes.
Rejections will be restricted to other energies, whose states are never
included into the averaging statistics. Let's consider that the maximum
energy jump allowed by the adopted protocol of (real) movements
corresponds to $n$ levels above or below the current energy $E$. Then,
let's take an energy window of $2n+1$ adjacent levels, starting from some
state inside it. Any movement which keeps the system still inside this
window will be accepted. This is the dynamic strategy we propose here.
Averages are taken only for the central level $E$ inside the chosen energy
window: the system is allowed to visit the other $n$ levels above it, as
well as the $n$ levels below it, nevertheless without measuring anything
during these side visits. Note that no tossed movement will be rejected,
if the current energy is just $E$. Thus, the averaging process is
completely rejection-free, avoiding any systematic bias due to artificial
rejections rules. The same strategy can be easily applied also for
continuous energy spectra, by taking averages only within a narrow,
rejection-free energy window centered inside another broader, free-visit
window: any movement to outside this latter would be rejected.

	In what follows, we consider a $L \times L$ square lattice Ising
ferromagnet, with $L = 32$ for which the exact function $g(E)$ is known
[21]. The energy of the current state $S$ is counted as the total number
of its unsatisfied bonds, i.e. the total number of neighbouring pairs of
spins pointing one up and the other down. The energy spectrum corresponds
to all even numbers between $0$ and $2,048$, i.e. $E = 0$, 2, 4, 6, 8
$\dots$ 2,048, which can also be represented by energy densities $e =
E/(2L^2)$. The degeneracies are $g(E) = 2$, 0, 2,048, 4,096, 1,057,792
$\dots$ 2, respectively. This spectrum is symmetric in relation to its
center at $E_{\rm max} = 1,024$ (or $e = 0.5$), where $g(E_{\rm max})
\approx 6.3 \times 10^{306}$. We need only its first half, corresponding
to positive temperatures. The critical energy (at the thermodynamic limit)
corresponds to $e_c \approx 0.147$ ($E_{\rm c} \approx 300$, for $L =
32$).

	We will adopt single-spin flips as the protocol of movements (both
for the real movements tossed during the computer run, as well as for the
virtual ones we consider in order to count $N^{\rm up}$ and $N^{\rm dn}$).
Starting from the current state $S$ with energy $E$, $1,024$ movements
would be available, by tossing one random spin to flip. They can be
classified according to the possible energy jumps, $E \to E\pm\Delta{E}$,
where $\Delta{E}$ could be $0, 2$ or $4$. Thus, our energy window will
have 5 adjacent levels, the central one, $E$, where the averages will be
measured, which is rejection-free, plus two neighbouring levels above it,
and two others below it. During the random walk performed inside this
window, we measure the values of $N^{\rm up}$ and $N^{\rm dn}$ for the
current state, every time its energy is $E$, and accumulate them into two
$E$-histograms $H^{\rm up}(E)$ and $H^{\rm dn}(E)$. Also a visit counter
$V(E)$ is upgraded to $V(E)+1$. If the energy of the current state is not
$E$, nothing is measured or accumulated. At the end, we take the averages
$<N^{\rm up}(E)> \,= H^{\rm up}(E)/V(E)$ and $<N^{\rm dn}(E)> \,= H^{\rm
dn}(E)/V(E)$. Note that this is the only role played by the final $V(E)$:
no comparison with the neighbouring values $V(E\pm\Delta{E})$ is needed,
no further information is extracted from these values. They must only be
large enough in order to provide a good statistics. For each new $E$, a
new 5-levels energy window centered on it is sampled, and the whole
process is repeated.

	By following this dynamic rule and by using the BHM relation (4),
we measured the quantity $\ln[g(E+\Delta{E})/g(E)]$ for 15 adjacent levels
$E = 300$, 302 $\dots$ 328, at the critical region. The deviations from
the exact values were averaged (root mean square) over these 15 levels,
and are shown as a function of the number of averaging states sampled
inside each energy level (counts), in figure 1. The squares corresponds to
$\Delta{E} = 4$, while the diamonds represent $\Delta{E} = 2$. The dashed
straight line ($1/\sqrt{counts}$) indicates that no systematic errors
besides the normal statistical fluctuations occur, giving credit to our
simple dynamic rule. According to these results, to improve the numerical
accuracy is a simple matter of taking more and more averaging states
inside each energy level, up to the computer time available.

	In order to perform thermal averages, one does not need the same
accuracy along the whole energy axis. The function $[g(E)\exp{(-E/Tc)}]^2$
is displayed by the dotted line in figure 2, where $T_c = 2.293930$ is the
exact location of the specific heat peak. This curve displays the
(squared) relative contribution of each energy to the partition function.
As the number of sampled averaging states inside each energy level is
proportional to the squared numerical accuracy, the dotted line in figure
2 shows the ideal profile of visits one needs in order to have equally
accurate contributions from each energy level. It is a sharp peaked curve,
according to which the computational effort can be concentrated only
inside a narrow energy window. Profiting from this feature, we have shaped
the profile of visits displayed by the solid line, in figure 2. The
possibility of designing this profile of visits, sampling different
numbers of averaging states for different energies, according to the
relative contribution of each energy region, is a further big advantage of
BHM over multicanonical methods. Almost all the computational effort is
concentrated near the peak.

	Figure 3 shows a detailed comparison of our simulational results
with the exact specific heat curve, near its peak. Being a derivative,
which corresponds to a mathematical procedure in which numerical accuracy
is strongly compromised, this quantity is a good standard for worst-case
comparisons, moreover near its peak. Nevertheless, the relative deviations
we obtained are compatible with the number ($counts = 8 \times 10^9$) of
averaged states per energy level, at the peak. Moreover, better yet
accuracies were obtained along {\bf all} the temperature range from 0 to
$\infty$, always by using {\bf the same} simulational data, i.e. the same
averaged values for the BHM quantities $N^{\rm up}$ and $N^{\rm dn}$
measured during {\bf a single} computer run.

	In short, we have tested a simple dynamics which is very efficient
in measuring microcanonical averages. It is essentially the same dynamics
as introduced in [20], for other purposes. Here, the aim is to measure the
microcanonical averages of some particular macroscopic quantities defined
within the broad histogram method [1-3]. Once one knows these averaged
quantities as functions of the energy, the method provides the energy
degeneracy function $g(E)$ through an exact relation. During the same
computer run, the microcanonical average $<Q(E)>$ of the quantity $Q$ of
interest is also measured. Then, once one knows $g(E)$ and $<Q(E)>$, the
canonical thermal average $<Q>_T$ can be determined by equation (2), for
{\bf any, continuously varying} temperature $T$, {\bf without resorting
again to further computer simulations}. According to our tests, the
current dynamic rule does not introduce any systematic averaging bias,
besides the normal statistical fluctuations which decay proportionally to
$1/\sqrt{counts}$. Thus, by applying this dynamic rule to BHM, to improve
more and more the numerical accuracy is a simple matter of increasing the
computer time. Among the further advantages of the microcanonical dynamics
tested here, we can quote: $i$) its implementation simplicity, without
detailed balance and other complications; $ii$) no movement-rejections at
all, within the averaging energy level; $iii$) the possibility to shape
the profile of visits along the energy axis, according to the desired
accuracy; $iv$) no randomness at all is used in order to decide to perform
or not the currently tossed movement [22]; $v$) short and non-periodic
waiting time between consecutive averaging states [23].

	I am indebted to Kim Doochul who warned me about reference [20],
reading which I have had the idea to apply its dynamic rules to the broad
histogram method. D. Stauffer was kind enough to perform a critical
reading of the manuscript. This work is partially supported by brazilian
agencies CAPES, CNPq and FAPERJ.

\vskip 30pt {\bf References}\parG

\item{[1]} P.M.C. de Oliveira, T.J.P. Penna and H.J. Herrmann, {\it Braz. 
J. Phys.} {\bf 26}, 677 (1996) (also in Cond-Mat 9610041).\par

\item{[2]} P.M.C. de Oliveira, {\it Eur. Phys. J.} {\bf B6}, 111 (1998) (also
in Cond-Mat 9807354).\par

\item{[3]} P.M.C. de Oliveira, {\it Braz. J. Phys.} {\bf 30}, 195 (2000)
(also in Cond-Mat 0003300).\par

\item{[4]} P.M.C. de Oliveira, T.J.P. Penna and H.J. Herrmann, {\it Eur. 
Phys. J.} {\bf B1}, 205 (1998); P.M.C. de Oliveira, in {\sl Computer
Simulation Studies in Condensed Matter Physics} {\bf XI}, 169, eds. D.P.
Landau and H.-B. Sch\"uttler, Springer, Heidelberg/Berlin (1998).\par

\item{[5]} P.M.C. de Oliveira, {\it Int. J. Mod. Phys.} {\bf C9}, 497
(1998).\par

\item{[6]} J.-S. Wang, T.K. Tay and R.H. Swendsen, {\it Phys. Rev.
Lett.} {\bf 82}, 476 (1999).\par

\item{[7]} J.-S. Wang, {\it Eur. Phys. J.} {\bf B8}, 287 (1999) (also
in Cond-Mat 9810017).\par

\item{[8]} J.D. Mu\~noz and H.J. Herrmann, {\it Int. J. Mod. Phys.} {\bf
C10}, 95 (1999); {\it Comput. Phys. Comm.} {\bf 121-122}, 13 (1999).\par

\item{[9]} R.H. Swendsen, B. Diggs, J.-S. Wang, S.-T. Li, C. Genovese and
J.B. Kadane, {\it Int. J. Mod. Phys.} {\bf C10}, 1563 (1999)  (also in
Cond-Mat 9908461).\par

\item{[10]} P.M.C. de Oliveira, {\it Comput. Phys. Comm.} {\bf
121-122}, 16 (1999), presented at the APS/EPS Conference on
Computational Physics, Granada, Spain (1998).\par

\item{[11]} J.-S. Wang, {\it Comput. Phys. Comm.} {\bf 121-122},
22 (1999).\par

\item{[12]} A.R. de Lima, P.M.C. de Oliveira and T.J.P. Penna, {\it
Solid State Comm.} {\bf 114}, 447 (2000) (also in Cond-Mat 9912152).\par

\item{[13]} J.-S. Wang and L.W. Lee, {\it Comput. Phys. Comm.} {\bf
127}, 131 (2000) (also in Cond-Mat 9903224).\par

\item{[14]} J.-S. Wang, {\it Prog. Theor. Phys. Supp.} {\bf 138}, 454
(2000) (also in Cond-Mat 9909177).\par

\item{[15]} A.R. de Lima, P.M.C. de Oliveira and T.J.P. Penna, {\it J.
Stat. Phys.} {\bf 99}, 691 (2000) (also in Cond-Mat 0002176).\par

\item{[16]} M. Kastner, J.D. Mu\~noz and M. Promberger, {\it Phys.
Rev.} {\bf E}, to appear (also in Cond-Mat 9906097).\par

\item{[17]} B.A. Berg and T. Neuhaus, {\it Phys. Lett.} {\bf B267}, 249
(1991); A.P. Lyubartsev, A.A. Martsinovski, S.V. Shevkunov and P.N.
Vorontsov-Velyaminov, {\it J. Chem. Phys.} {\bf 96}, 1776 (1992); E.
Marinari and G. Parisi, {\it Europhys. Lett.} {\bf 19}, 451 (1992); B.A.
Berg, {\it Int. J. Mod. Phys.} {\bf C4}, 249 (1993).\par

\item{[18]} J. Lee, {\it Phys. Rev. Lett.} {\bf 71}, 211 (1993).\par

\item{[19]} B. Hesselbo and R.B. Stinchcombe, {\it Phys. Rev. Lett.}
{\bf 74}, 2151 (1995).\par

\item{[20]} K.-C. Lee, {\it J. Phys.} {\bf A23}, 2087 (1990).\par

\item{[21]} P.D. Beale, {\it Phys. Rev. Lett.} {\bf 76}, 78 (1996).\par

\item{[22]} Within single-spin flips, for instance, random numbers are
used only to toss which spin would be flipped, a coarse grained decision
concerning only integer numbers. One does not need to compare them with
precisely defined acceptance ratios, a delicate comparison between {\bf
real} numbers which could introduce undesired biases as in\par

\item{} A.M. Ferrenberg, D.P. Landau and Y.J. Wong, {\it Phys. Rev.
Lett.} {\bf 69}, 3382 (1992).\par

\item{[23]} In the present case of a $32 \times 32$ lattice, a new state
was sampled after every $\sim 40$ spin flips, in average, instead of $L^2
= 1,024$.\par

\vskip 30pt {\bf Figure Captions}\parG

\item{1} Deviations between measured $g(E)$ and the exact values, as a
function of the number ($counts$) of sampled averaging states. The dashed
line corresponds to $1/\sqrt{counts}$. Data for $32 \times 32$ Ising
ferromagnet.\par

\item{2} Profile of visits along the energy axis, which could be shaped
according to the importance of each energy contribution to the partition
function (dotted line).\par

\item{3} Detail of the specific heat peak, a worst-case comparison.\par

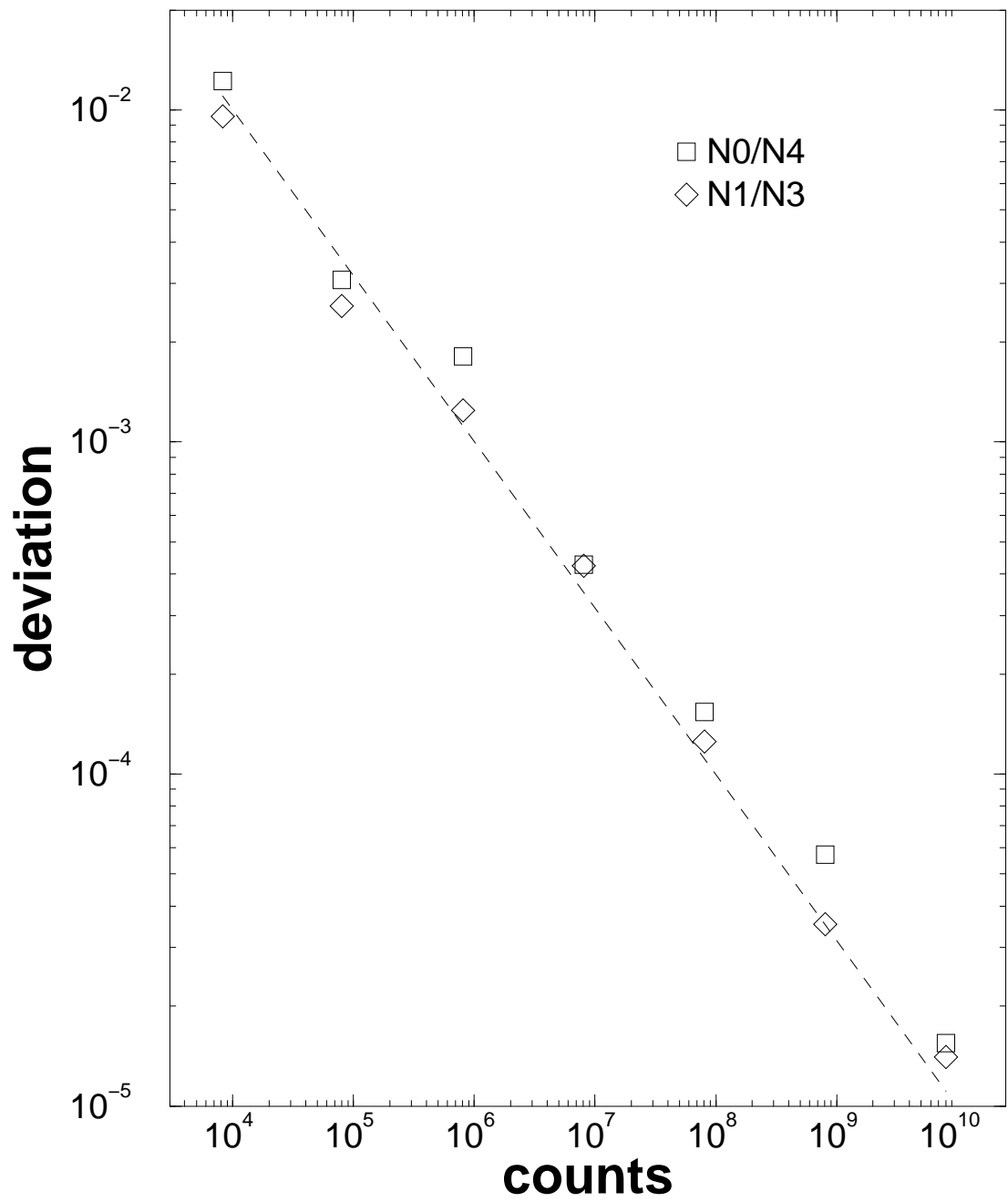
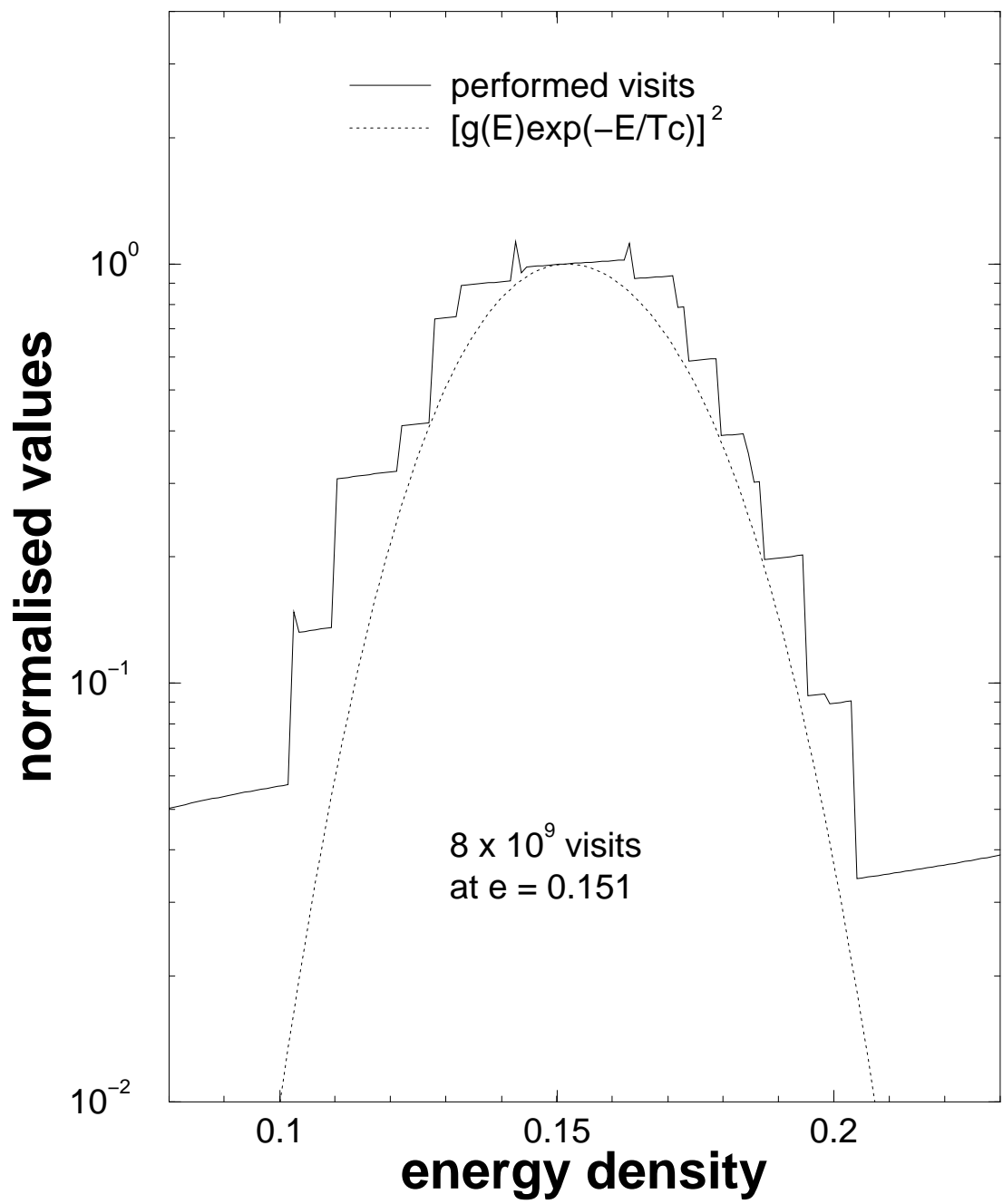
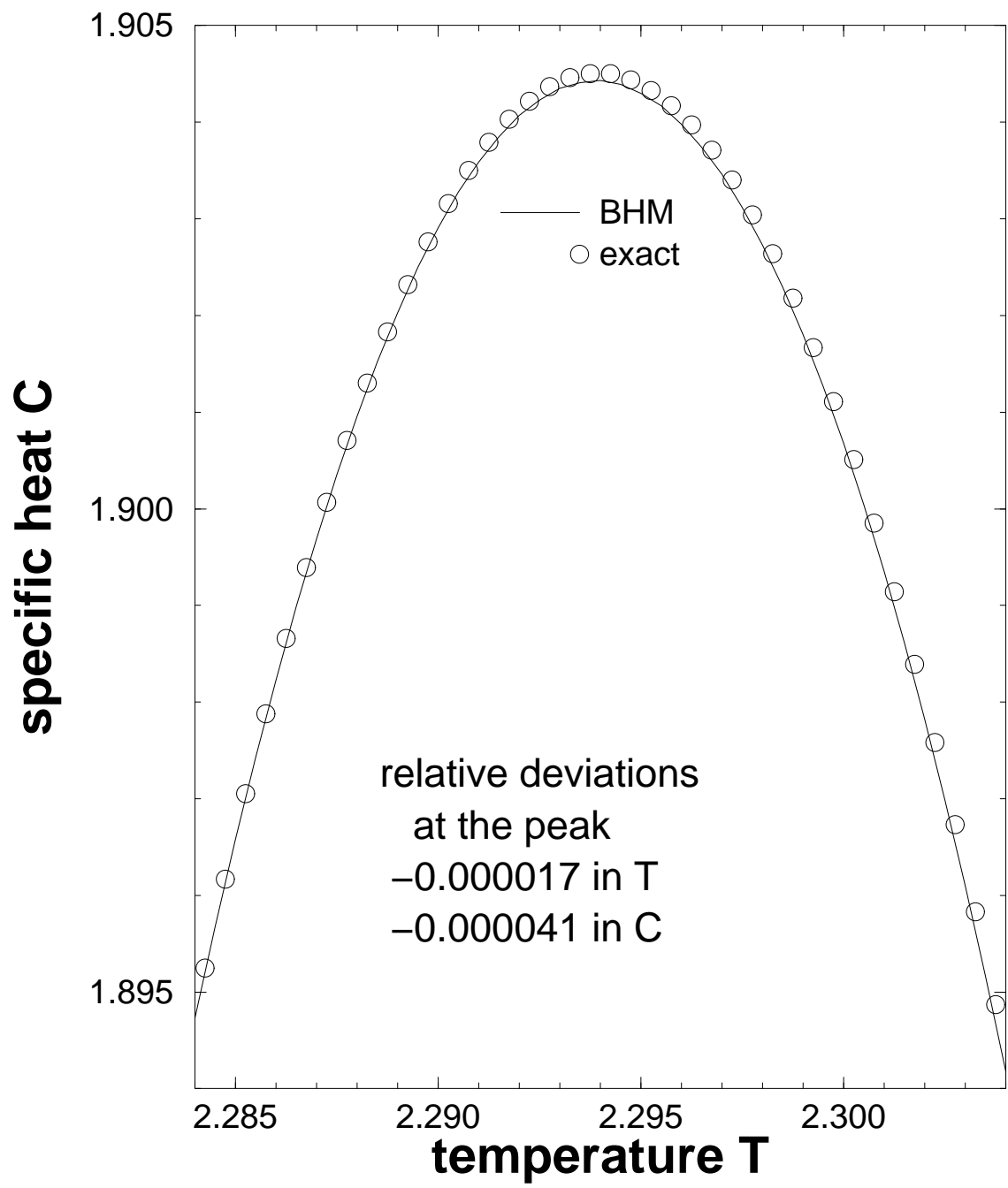
\bye